\documentclass{article}
\usepackage{spconf,amsmath,graphicx}
\usepackage{url}

\title{Soundata: A Python library for reproducible use of audio datasets}
\name{$\begin{array}{cc} \mbox{ Magdalena Fuentes$^{1*}$,  Justin Salamon$^{2*}$, Pablo Zinemanas$^{3}$, Mart\'in Rocamora$^{4}$,} \\
\mbox{Gen\'is Paja$^{3}$, Ir\'an R. Rom\'an$^{1}$,  Marius Miron$^{3}$, Xavier Serra$^{3}$, Juan Pablo Bello$^1$ } \end{array}$ }

\address{$^1$ New York University, New York, United States\\
$^2$ Adobe Research, San Francisco, United States\\
$^3$ Universitat Pompeu Fabra, Barcelona, Spain\\
$^4$ Universidad de la Rep\'ublica, Montevideo, Uruguay\\
{\normalsize $^*$Equal contribution}}
\begin{document}
%
\maketitle
\begin{abstract}
Soundata is a Python library for loading and working with audio datasets in a standardized way, removing the need for writing custom loaders in every project, and improving reproducibility by providing tools to validate data against a canonical version. It speeds up research pipelines by allowing users to quickly download a dataset, load it into memory in a standardized and reproducible way, validate that the dataset is complete and correct, and more. Soundata is based and inspired on \textit{mirdata} \cite{bittner2019mirdata}, and design to complement mirdata by working with environmental sound, bioacoustic and speech datasets, among others. Soundata was created to be easy to use, easy to contribute to, and to increase reproducibility and  standardize usage of sound datasets in a flexible way.
\end{abstract}
\section{Statement of need}

As research pipelines become increasingly complex, it is key that their different components are reproducible. 
In recent years, the machine listening research community has done considerable efforts towards standardization and reproducibility, by using common libraries for modelling and evaluation \cite{tensorflow, pedregosa2011scikit, chollet2015keras, mesaros2016metrics}, open sourcing models \cite{zinemanas2020dcase, speechbrain} and data using resouces such as Zenodo\footnote{\url{https://zenodo.org}}. However, it has been previously shown that discrepancies in local version of the data and different practices in loading and parsing datasets can lead to considerable differences in performance results, which is misleading when comparing methods \cite{bittner2019mirdata}. Besides, from a practical point of view, it is extremely inefficient to develop pipelines from scratch for loading and parsing a dataset each researcher or team each time, and increases the chances of bugs in the code and differences that hinders reproducibility.

Mirdata \cite{bittner2019mirdata} was introduce as a tool for mitigating the lack of reproducibility and efficiency when working with datasets in the context of Music Information Research (MIR). In MIR, the mentioned issues are exacerbated due to the intrinsic commercial nature of music data, since it is very difficult to get licenses to distribute music recordings openly. Mirdata has grown considerably in the past year with the addition of multiple dataset loaders and contributions from the community. 

Musical datasets are extremely complex compared to other audio datasets. They usually convey much more metadata information (e.g. artists, musicians, instruments, etc.) and their annotations are 
of several different types and formats, reflecting the several different tasks that there are in Music Information Research (e.g melody estimation, beat tracking, chord estimation, among others). Dedicating the same repository for music and other purpose audio datasets would converge to a very complex, hard to manage repository, which would be difficult to scale. Instead, we introduce Soundata\footnote{\url{https://github.com/soundata/soundata}} as a separated package which is inspired and based on Mirdata, but deals with the annotation types and formats that the communities working with environmental sound, bioacoustics, and speech would require. 
 
There are other libraries that handle datasets, like \cite{tensorflow, zinemanas2020dcase}. However, using datasets in the context of those libraries makes it difficult to interchange models and software, as data loaders are designed to work with those environments and further adaptation is required to migrate them. Instead, we think that data should be handled separately, as a standalone library that can easily be plugged in different work pipelines, with different modelling software.

Soundata is based and inspired on \textit{mirdata} \cite{bittner2019mirdata}, and was created following these design principles:\\
\textbf{Easy to use:} Soundata is designed to be easy to use and to simplify the research pipeline considerably. Check out the examples in the Getting started page.\\
\textbf{Easy to contribute to:} we welcome and encourage contributions, especially new datasets. You can contribute following the instructions in our Contributing page.\\
\textbf{Increase reproducibility:} by providing a common framework for researchers to compare and validate their data, when mistakes are found in annotations or audio versions change, using Soundata the audio community can fix mistakes while still being able to compare methods moving forward.\\
\textbf{Standardize usage of sound datasets:} we standardize common attributes of sound datasets such as audio or tags to simplify audio research pipelines, while preserving each dataset’s idiosyncrasies: if a dataset has ‘non-standard’ attributes, we include them as well.


\bibliographystyle{IEEEbib}
\bibliography{strings,refs}

\end{document}